\documentclass[longbibliography, reprint, amsmath,amssymb, aps, prl, superscriptaddress, nofootinbib]{revtex4-2}

\usepackage{graphicx}
\usepackage{dcolumn}
\usepackage{bm}
\usepackage{physics}
\usepackage{hyperref}
\usepackage{orcidlink}

\begin{document}

\title{Spectroscopy of ultracold molecules in the mid-infrared}

\author{J. Rodewald\orcidlink{0009-0006-5794-1890}}
\affiliation{Centre for Cold Matter, Blackett Laboratory, Imperial College London, London SW7 2AZ, United Kingdom}
\author{J. R. Cox \orcidlink{0009-0000-5447-1471}}
\affiliation{Centre for Cold Matter, Blackett Laboratory, Imperial College London, London SW7 2AZ, United Kingdom}
\affiliation{Quantum Engineering Centre for Doctoral Training, University of Bristol, Bristol BS8 1FD, United Kingdom}
\author{J. Vagge}
\affiliation{Centre for Cold Matter, Blackett Laboratory, Imperial College London, London SW7 2AZ, United Kingdom}
\author{Q. Lyu\orcidlink{0009-0007-6310-5714}}
\affiliation{Centre for Cold Matter, Blackett Laboratory, Imperial College London, London SW7 2AZ, United Kingdom}
\author{K. K. Voges\orcidlink{0000-0003-3773-4270}}
\altaffiliation[Current address: ]{Institut für Quantenoptik, Leibniz Universität Hannover, 30167 Hannover, Germany}
\author{B. E. Sauer\orcidlink{0000-0002-3286-4853}}
\affiliation{Centre for Cold Matter, Blackett Laboratory, Imperial College London, London SW7 2AZ, United Kingdom}
\author{F. Wiotte\orcidlink{0009-0009-8355-9865}}
\affiliation{Laboratoire de Physique des Lasers, CNRS, Universit\'e Sorbonne Paris Nord, 93430 Villetaneuse, France}
\author{A. Bonifacio\orcidlink{0009-0006-7313-8176}}
\affiliation{Laboratoire de Physique des Lasers, CNRS, Universit\'e Sorbonne Paris Nord, 93430 Villetaneuse, France}
\author{M. Manceau\orcidlink{0000-0002-8319-5741}}
\affiliation{Laboratoire de Physique des Lasers, CNRS, Universit\'e Sorbonne Paris Nord, 93430 Villetaneuse, France}
\author{B. Darqui\'e\orcidlink{0000-0002-8623-5817}}
\affiliation{Laboratoire de Physique des Lasers, CNRS, Universit\'e Sorbonne Paris Nord, 93430 Villetaneuse, France}
\author{M. R. Tarbutt\orcidlink{0000-0003-2713-9531}}\email{m.tarbutt@imperial.ac.uk}
\affiliation{Centre for Cold Matter, Blackett Laboratory, Imperial College London, London SW7 2AZ, United Kingdom}

\begin{abstract}
Unlike the microwave and optical domains, frequency metrology in the mid-infrared has not yet benefited from the long coherence times and exquisite control provided by ultracold particles. We use a quantum cascade laser operating at 17.5~THz to drive ro-vibrational transitions in ultracold calcium monofluoride molecules. The laser linewidth is narrowed to 15~kHz by locking it to a custom mid-infrared ring cavity, which is in turn locked to an ultrastable laser. We study the spectrum for molecules confined in red-detuned and blue-detuned magneto-optical traps, and in an optical dipole trap. In the dipole trap, the spectroscopic linewidth is close to that of the laser. We consider the stability and accuracy that can be reached using these methods in the context of a broader programme for improving frequency metrology in the mid-infrared. 

\end{abstract}

\maketitle

Spectroscopy in the mid-infrared (MIR), known as the molecular fingerprint region, is an important and growing technology. It is useful whenever highly selective and sensitive molecular identification and quantitative analysis are needed. This makes it a powerful tool for trace gas detection~\cite{Kosterev2008} with applications in breath analysis~\cite{Henderson2018}, environmental monitoring~\cite{Li2022, Tian2023, Teng2026}, atmospheric physics~\cite{Galli2016, Daghestani2014, Robinson2021}, and astrophysics~\cite{Hernan-Caballero2011, Najita2021}. In molecular physics, MIR spectroscopy is used to determine molecular constants~\cite{Asselin2017, Wang2025, Tran2025, Asselin2026arxiv}, validate quantum chemistry calculations~\cite{Polyansky2012}, improve spectroscopic databases~\cite{Gordon2022, NIST118, Tennyson2024} and study the fundamental interactions of molecules with surfaces~\cite{LukusaMudiayi2021, Arellano2024}. Sensitivity at the single molecule level can be reached using quantum logic or action spectroscopy techniques~\cite{Calvin2025, Shlykov2026arxiv}. Quantum state resolved MIR spectroscopy of large molecules, such as fullerenes, can be used to study the dynamics of strongly-correlated many-body systems~\cite{Shulz2015, Changala2019, Liu2022, Liu2023b}. MIR frequency measurements are also important for improving the determination of fundamental constants~\cite{Mejri2015, Alighanbari2020, Patra2020} and for testing new ideas in fundamental physics. Measurements of molecular vibrational frequencies are sensitive to variations in the electron-to-proton mass ratio ($m_e/m_p$) predicted by unification theories and some theories of dark matter and dark energy~\cite{Barontini2022Short, Leung2022, Chou2017, Sinhal2020, Elder2025}.  With improved MIR frequency metrology, it should be possible to measure the parity-violating (PV) energy difference between left- and right-handed chiral molecules, which will be an important measure of electroweak interactions in normal matter, and sensitive to new PV interactions beyond the Standard Model~\cite{Cournol2019, Fiechter2022}. For some molecules, the predicted energy difference is a few Hz, and the vibrational modes are close to the MIR frequency used in the present work~\cite{Eduardus2026, Eduardus2026b}. These applications call for ultrastable lasers and accurate frequency standards -- the tools of frequency metrology -- to be extended into the MIR domain.

Frequency metrology in the microwave and optical domains has been transformed by the availability of ultracold atoms and ions, and of ultrastable lasers and frequency combs. Trapped, ultracold particles enable long coherence times, elimination of Doppler shifts, strong spatial localization, and outstanding control over frequency shifts. This transformation has not yet extended to the MIR because atoms and ions do not usually have transitions in this region, and because MIR laser technology is not as advanced. There are only two recommended frequency standards throughout this region, at 3.39~$\mu$m and 10.3~$\mu$m, and these are $10^3 - 10^4$ times less accurate than microwave and optical standards~\cite{Quinn2003}. Two recent advances raise the prospect of far better MIR frequency metrology. First, a variety of molecules have been cooled to ultracold temperatures and trapped~\cite{Fitch2021b, Augenbraun2023, Langen2024, Deiss2023}, offering many MIR transitions that could be used as standards.  Second, quantum cascade lasers (QCLs) are becoming available throughout the MIR~\cite{Van2019, Razeghi2025}, and metrology-level frequency stabilization of these lasers has been demonstrated~\cite{Argence2015}. Here, we bring these techniques together to demonstrate vibrational spectroscopy of ultracold molecules with a MIR laser.

\begin{figure}[t]
    \centering
    \includegraphics[width=\columnwidth]{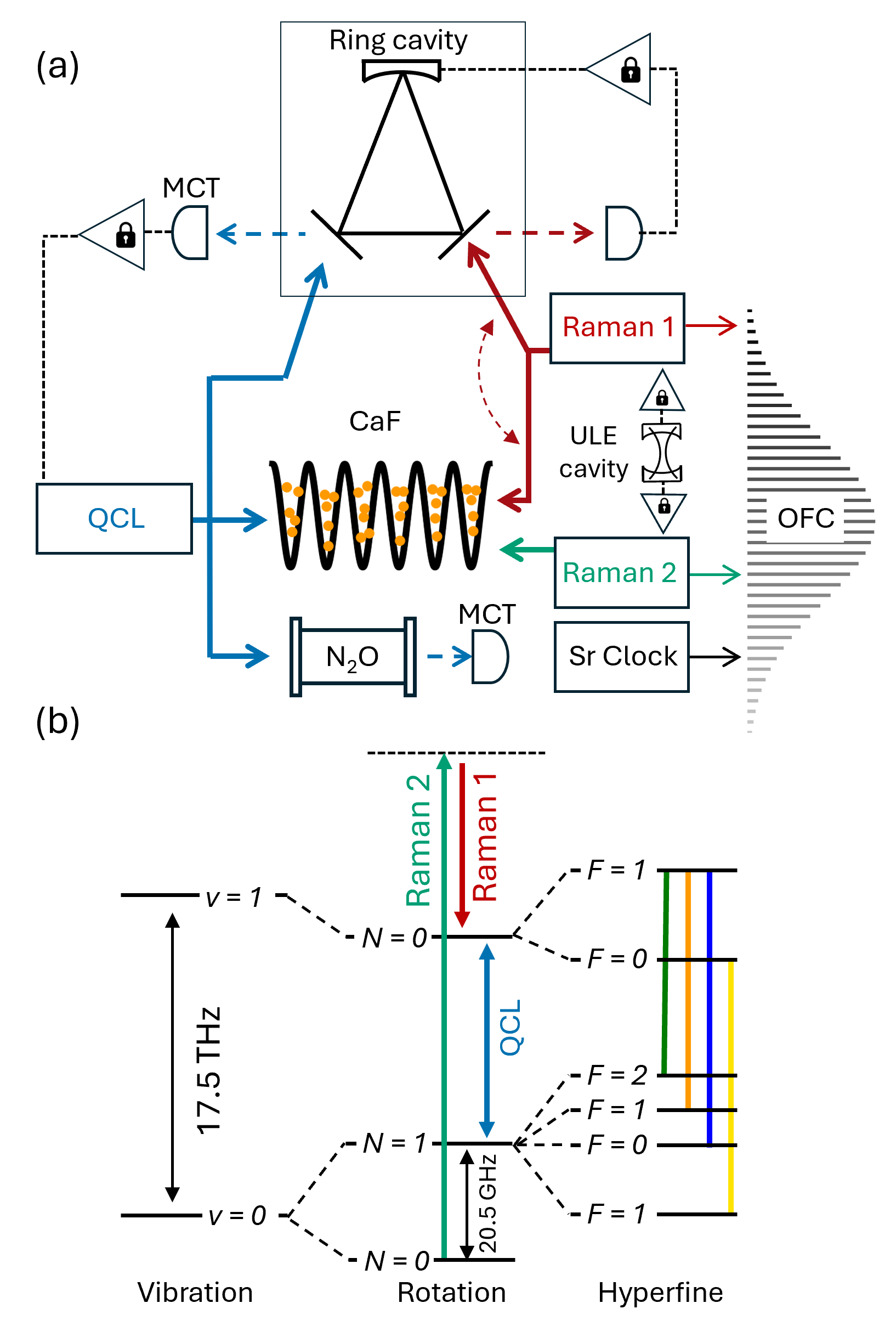}
    \caption{MIR frequency metrology with ultracold CaF in a dipole trap or optical lattice. (a) The vibrational transition can be driven using two photons from a pair of ultrastable near infrared lasers locked to an ultra-low expansion (ULE) cavity and measured using an optical frequency comb (OFC). It can also be driven directly with a quantum cascade laser (QCL) at 17.5~THz, as demonstrated here. The QCL is locked to a ring cavity which is in turn locked to one of the ultrastable lasers. The same QCL is used for spectroscopy of N$_2$O. The MIR power is measured using mercury cadmium telluride (MCT) detectors. (b) Relevant energy levels of CaF~\cite{Anderson1994, Childs1981}. }
    \label{fig:scheme}
\end{figure}

Figure \ref{fig:scheme} illustrates our setup for probing the 17.5~THz vibrational transition of trapped, ultracold CaF molecules using both one-photon and two-photon spectroscopy.  The pure vibrational frequency can be measured by driving the Raman transition $\ket{v,N} = \ket{0,0} \leftrightarrow \ket{1,0}$\footnote{Throughout, we label states as $\ket{v,N,F}$, but omit $F$ when we are referring to all hyperfine components.} using a pair of lasers locked to an ultra-stable cavity and measured with a frequency comb referenced to a local Sr lattice clock~\cite{Baynham2026}. This transition has a natural linewidth below 1~Hz, and previous analyses suggest that an accuracy of 10$^{-17}$ can be reached~\cite{Kajita2018, Barontini2022Short}. The rotational frequency can be measured to a similar absolute accuracy by microwave spectroscopy~\cite{Williams2018, Blackmore2018}, establishing the frequency of the $\ket{0,1} \leftrightarrow \ket{1,0}$ transition. The focus of the present work is to probe this latter transition.

\begin{figure}[t]
    \centering
    \includegraphics[width=\columnwidth]{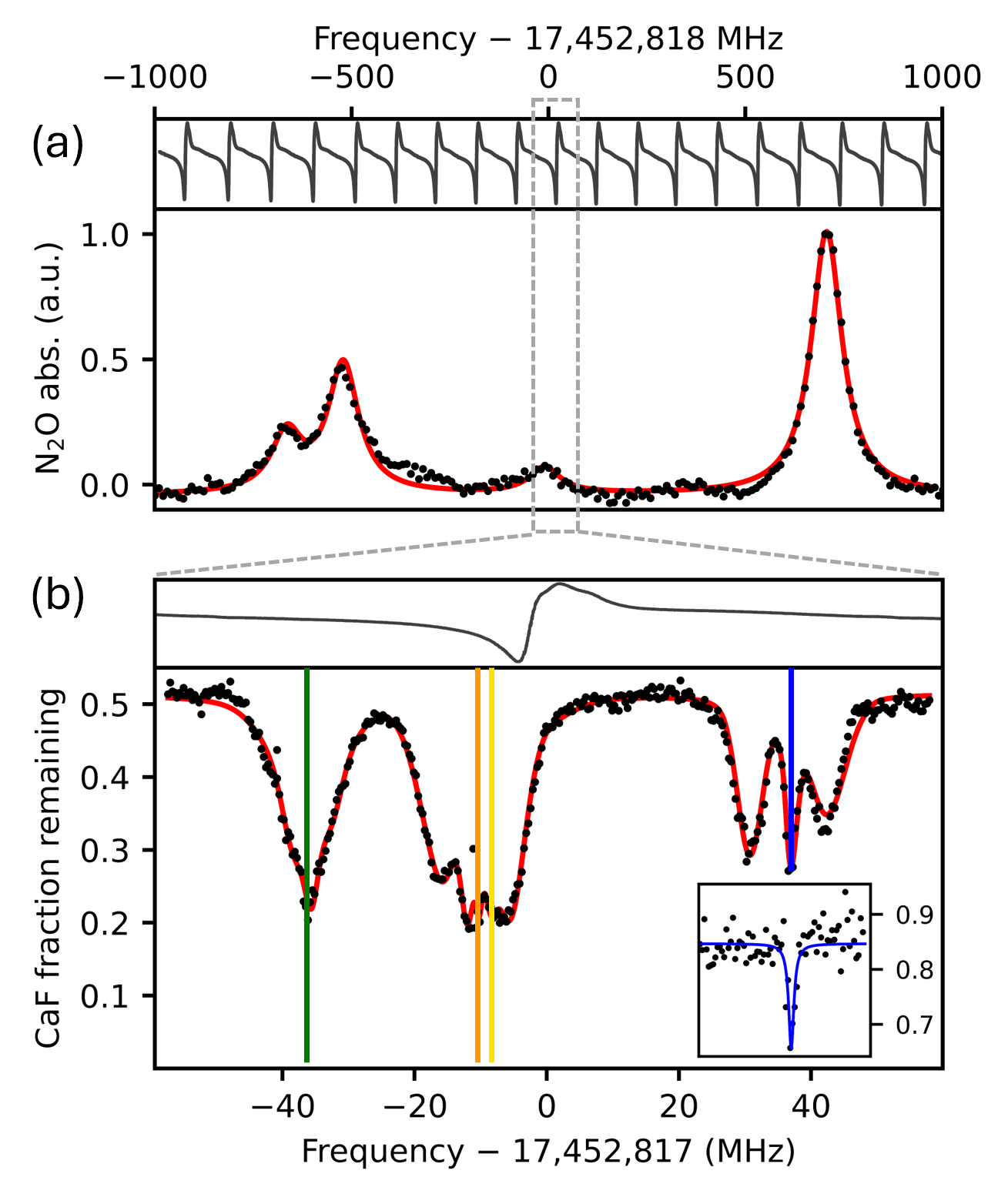}
    \caption{(a) N$_2$O absorption spectrum used to determine the QCL frequency; there is a Q branch line on the right and two overlapping R branch lines on the left. The optical depth is small. Points: data. Line: Fit to a sum of Lorentzians with amplitudes as free parameters and frequencies fixed to those given in \cite{Wang2025}; these fix the absolute frequency to about 3~MHz, limited by the accuracy of the N$_2$O lines. Top: cavity fringes used to lock the QCL. (b) MIR spectroscopy of the $\ket{0,1}\leftrightarrow\ket{1,0}$ transition of CaF in a MOT. Coloured vertical lines correspond to components indicated in Fig.~\ref{fig:scheme}(b). Zeeman splitting and broadening is evident. Points: data. Line: model discussed in text. Inset: The $\ket{0,1,0}\leftrightarrow \ket{1,0,1}$ transition for molecules in a blue-detuned MOT. Top: cavity fringe.}
    \label{fig:MOTSpectrum}
\end{figure}

The 17.5~THz laser is a continuous wave distributed feedback QCL based on an InAs/AlSb material~\cite{Manceau2026}. It provides up to 1~mW of power and its frequency is tuneable by about 100~GHz. The injection current is digitally controlled with a resolution of 1~$\mu$A, corresponding to a frequency resolution of $\sim250$~kHz. Fast analog current modulation and feedback is via a shunt path that diverts current to and from the QCL without perturbing the driver. There being no wavemeters in this spectral region, we use absorption spectroscopy of N$_2$O to determine the laser frequency within a few MHz~\cite{Wang2025}. Figure \ref{fig:MOTSpectrum}(a) shows the absorption spectrum of the bending mode of N$_2$O near the frequency of the CaF transition. Here, the frequency is controlled by scanning the QCL current directly. While this works well for the broad N$_2$O lines, scanning this way yields non-reproducible CaF spectra where the lines are much narrower. Thus, an important part of the present work was to implement a reliable frequency scanning method, which we now describe.

The QCL output is split into two equal parts, with one half sent either to CaF or to N$_2$O, and the other to a ring cavity used to lock the laser frequency. The cavity has a free spectral range of 104~MHz and a MIR finesse of 16 (see Appendices A and B). The upper part of Fig.~\ref{fig:MOTSpectrum}(a) shows the cavity signal as the QCL frequency is scanned. The laser frequency and cavity length are not modulated, yet the cavity resonances have a dispersive shape. We explain this in Appendix B. With a DC offset subtracted these features provide steep zero crossings convenient for laser locking, which is done by feeding back to the QCL current shunt path with a bandwidth of 50~kHz. The cavity length is locked to a reference laser at 1134~nm (Raman 1 in Fig.~\ref{fig:scheme}(a)) using the Pound-Drever-Hall (PDH) method. Feedback is to a stack of two piezo actuators which control the position of the curved mirror with a locking bandwidth of $\sim$300~Hz. The reference laser can be locked to a high precision wavemeter, providing wide tuneability and a root mean square (RMS) stability of 200~kHz at 1~s. For higher stability and finer frequency scanning, we add sidebands to the laser with an electro-optic modulator (EOM) and lock a sideband to an ultra-low expansion (ULE) high-finesse cavity using the PDH method. This yields a minimum Allan deviation of 2.5~Hz at an averaging time of 20~s, measured via heterodyne beat with an optically-stabilized frequency comb. Scanning the EOM scans the reference laser, which then scans the length of the ring cavity and thus the QCL frequency. We trigger an automatic re-lock routine when the ULE cavity transmission drops below a threshold so that the reference laser can be scanned in steps larger than the cavity linewidth. These methods provide sub-Hz tuning resolution of the QCL frequency and narrow its linewidth from 167~kHz when free-running to 15.2~kHz. Further details of the locking scheme and measurement of the laser linewidth are in Appendix A.

We capture and cool about $5 \times 10^4$ CaF molecules in a red-detuned magneto-optical trap (MOT), following the procedures described previously~\cite{Truppe2017, Truppe2017b, Williams2017, Williams2018, Caldwell2019}. The intensity of the MIR light at the molecules, $I$, is controlled by tuning the QCL temperature and using germanium attenuators. The interaction time, $\tau$, is controlled using a mechanical shutter with a rise time of 1~ms. As a first experiment, we use the QCL to drive the vibrational transition of molecules trapped in the MOT with a temperature $T\approx 10$~mK~\cite{Williams2017}. We take a first image of the MOT, turn on the QCL light for $\tau=60$~ms, then take a second image. Figure \ref{fig:MOTSpectrum}(b) shows the fraction of molecules remaining as a function of the QCL frequency. The MIR light introduces loss from the MOT when it is resonant with any of the six hyperfine components of $\ket{0,1} \leftrightarrow \ket{1,0}$. Four of the six lie within the range scanned here, centered at the colored vertical lines. In a separate experiment, we determine the frequencies of these lines by hyperfine-resolved laser-induced fluorescence spectroscopy of the P(1) components of $A{}^{2}\Pi_{1/2}(v=0) \leftarrow X{}^{2}\Sigma^+(v=0,1)$ using a laser crossing a CaF beam at 90$^{\circ}$. The laser frequencies are measured using a wavemeter calibrated to the Rb D2 line. Their difference, together with the known rotational and hyperfine frequencies~\cite{Anderson1994, Childs1981}, establishes the frequency of $\ket{0,1}\leftrightarrow \ket{1,0}$ with 1.2~MHz uncertainty. 

In red-detuned molecular MOTs, polarization gradient heating dominates at low magnetic field ($B$) and low velocity ($v$), while Doppler cooling dominates at higher $B$ and $v$~\cite{Devlin2016, Devlin2018}. As a result, these MOTs have a temperature well above the Doppler limit and are displaced from $B=0$. The resulting Zeeman broadening and splitting of the transitions is clearly visible in Fig.~\ref{fig:MOTSpectrum}(b), especially for the $\ket{0,1,0}\leftrightarrow \ket{1,0,1}$ line (blue line in Figs.~\ref{fig:scheme}(b) and \ref{fig:MOTSpectrum}(b)) which is split into three Zeeman components with the $m_F = \pm 1$ components broader than the central $m_F=0$ component. Using the known field gradient and Zeeman shifts of the hyperfine levels~\cite{Fitch2021b}, we calculate the distribution of Zeeman shifts for all the transitions in Fig.~\ref{fig:MOTSpectrum}(b), assuming a MOT centered at $\{x_0,y_0,z_0\}$ with an isotropic Gaussian distribution of RMS width $\sigma$, and use this to model the data. We fix $\{x_0,z_0\} = \{1.2,1.3\}$~mm, the values measured from fluorescence images of the MOT, and fix the hyperfine intervals to their known values~\cite{Childs1981}, but leave all amplitudes as free fit parameters. An additional source of broadening arises because the MIR light probes $\ket{v,N}$ dressed by the MOT light. The resulting Autler-Townes splitting is unresolved but leads to lines that are approximately Lorentzian with a width of $\gamma_{\rm eff} \approx \Gamma_{\rm A} \times (\Omega_{\rm MOT}/\Gamma_{\rm A})^2/(1+4\Delta_{\rm MOT}^{2}/\Gamma_{\rm A}^2)$, where $\Omega_{\rm MOT}$ and $\Delta_{\rm MOT}$ are the MOT Rabi frequency and detuning, and $\Gamma_{\rm A}$ is the decay rate of the excited state. The MOT has $\Omega_{\rm MOT} \approx \Delta_{\rm MOT} \approx \Gamma_{\rm A}$, so we expect $\gamma_{\rm eff} \approx \Gamma_{\rm A}/5 \approx 2$~MHz. Since $\Omega_{\rm MOT}$ and $\Delta_{\rm MOT}$ differ for each component, we allow a different $\gamma_{\rm eff}$ for each of the three features in the spectrum. The red line in Fig.~\ref{fig:MOTSpectrum}(b) shows the resulting fit, which gives $y_0 = 0.31$~mm, $\sigma = 0.52$~mm, and $\gamma_{\rm eff}=\{1.34, 3.28, 1.58\}$~MHz from left to right, all close to the above estimate. 

\begin{figure*}
    \centering
    \includegraphics[width=1\linewidth]{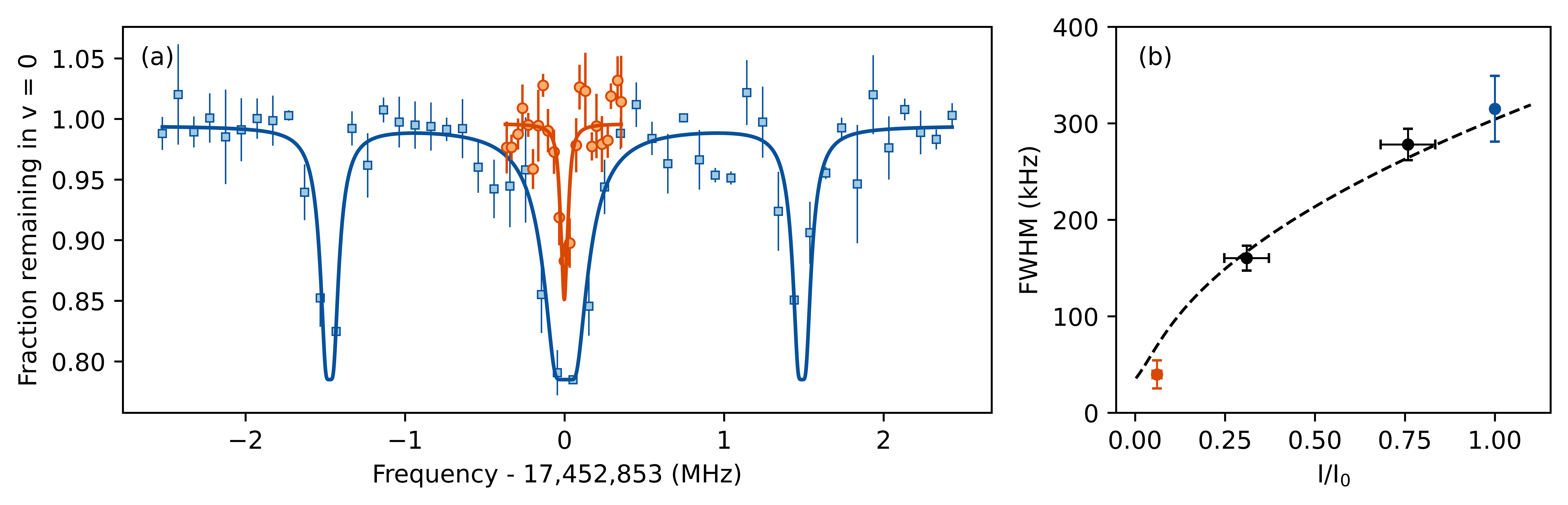}
    \caption{(a) Spectrum of the $\ket{0,1,0}-\ket{1,0,1}$ line of CaF in an optical dipole trap. Blue points: $I=I_0$, $B=1.05$~G. Orange points: $I=0.06I_0$, $B=0.0147$~G. Blue and orange lines: fits to lineshape model described in Appendix C. (b) FWHM for 4 different values of $I/I_0$. Dashed line: fit to Eq.~(\ref{eq:fwhm}). Horizontal error bars are uncertainties of relative power measurements.}
    \label{fig:ODT_Spectrum}
\end{figure*}

We repeat this spectroscopy in a blue-detuned MOT (BDM)~\cite{Jarvis2018}, where polarization gradient cooling dominates and the cloud is centered at $B=0$ and shrinks to $\sigma < 100$~$\mu$m. To load the BDM~\cite{Yu2026}, we first compress the red-detuned MOT~\cite{Williams2018}, then cool the cloud to 30~$\mu$K using $\Lambda$-enhanced gray molasses~\cite{Cheuk2018} at zero field, and finally ramp the field gradient up to 42~G/cm over 50~ms. We take a first image of the BDM, apply the QCL light for $\tau=50$~ms, then take a second image. The inset in Fig.~\ref{fig:MOTSpectrum} shows a scan over $\ket{0,1,0}\leftrightarrow \ket{1,0,1}$. We no longer observe any Zeeman splitting. Instead, the triplet reduces to a single line which fits well to a Lorentzian with a full width at half maximum (FWHM) of $0.96(27)$~MHz. This width is consistent with the small cloud size and the reduced scattering rate in the BDM where both the cooling and trapping rely on the presence of transient dark states~\cite{Lyu2026}. The higher baseline for the BDM data is due to the shorter time interval between the two images.

 Finally, we perform spectroscopy of the molecules in an optical dipole trap (ODT). The ODT has a wavelength of 1064~nm, a power of 18~W, and a waist of 33~$\mu$m, yielding a trap depth of 434~$\mu$K. We follow the same procedure as above, turning on the ODT at the start of the BDM field ramp. About 15\% of the molecules initially captured in the MOT are transferred to the ODT. After loading, we turn off the field gradient, cool the molecules to $T=40$~$\mu$K using a 25~ms pulse of $\Lambda$-molasses, wait for 5~ms for untrapped molecules to fall away, then image the molecules in the ODT with a 10~ms pulse of $\Lambda$-molasses light. We optically pump around half the molecules into $\ket{0,1,0}$ by applying 20~$\mu$W of light resonant with all hyperfine components except for the target state~\cite{Williams2018}. The molecules are held in the ODT for 50~ms while a uniform field $B$ turns on, and are then probed by the QCL for $\tau=4$~ms. Molecules in $\ket{0,1}$ are returned to the BDM where they are imaged to determine the fraction remaining in $v=0$ as a function of the QCL frequency. 
 
The blue points in Fig.~\ref{fig:ODT_Spectrum}(a) show the three $m_F$ components of $\ket{0,1,0}-\ket{1,0,1}$ measured at $I=I_0$ and $B=1.05(1)$~G. Here, $I_0$ is a reference intensity that we estimate below. The dips saturate near 75\% because about 50\% of the $v=0$ molecules are in $F=0$ and at most 50\% of those can be transferred to $v=1$. The blue line in Fig.~\ref{fig:ODT_Spectrum}(a) is a fit to the lineshape model described in Appendix C which accounts for both saturation broadening and the laser linewidth $\Gamma_{\rm L}$. We do a simultaneous fit to four datasets acquired at four different values of $I$, with $\Gamma_{\rm L}$ as a common free parameter. We assume the light is linearly polarized so that $m_F=\pm 1$ are driven equally. Since the polarization direction relative to $B$ is unknown, we take the intensity ratio of $\pi$ and $\sigma$ light as a global fit parameter. The intensity for each dataset is also a free fit parameter, because the measured values are uncertain. The model fits well to all four datasets, with frequency intervals consistent with the quadratic Zeeman shift of $m_F=0$, which we calculate to be 15.3~kHz/G$^2$. The orange points in Fig.~\ref{fig:ODT_Spectrum}(a) show the same line scanned at a much lower intensity, $I=0.06I_0$. Here, we attempted to set $B$ to zero, and measured $B=14.7$~mG using microwave spectroscopy of the rotational transition~\cite{Williams2018}. We fix the Zeeman splitting to the values calculated for this $B$. The fits determine the centre frequency with an uncertainty of 18.5~kHz for the blue data, and 4.5~kHz for the orange data. The global fit gives $\Gamma_{L} = 35 (14)$~kHz, consistent with the 15.2~kHz determined from the noise measurements presented in Appendix A. For our $T=40$~$\mu$K, the Doppler broadening is 10.3~kHz. 

Figure \ref{fig:ODT_Spectrum}(b) shows the width of the $m_F=0$ component for four values of $I$. The dashed line is a fit to Eq.~(\ref{eq:fwhm}) which gives $I_0 = 27(2)$~mW~m$^{-2}$. By measuring the power and divergence of the beam exiting the QCL, assuming it behaves as an ideal Gaussian beam, and estimating the loss at each optic, we estimate $I_0 = 15-230$~~mW~m$^{-2}$. The range is large because of the uncertainties in measuring the power and beam size, but overlaps with the value determined from the model.

It is interesting to consider the precision and accuracy that could be achieved in the future. In the present work, center frequencies are determined with a statistical uncertainty of a few kHz, but the absolute frequency was determined from the difference between two optical transitions measured with a wavemeter, limiting the accuracy to 1~MHz. We are currently working to drive $\ket{0,0}-\ket{1,0}$ using the two Raman lasers illustrated in Fig.~\ref{fig:scheme}(a) which are locked to an ultrastable cavity and referenced to a frequency comb.  Previous work~\cite{Kajita2018, Barontini2022Short} estimates that a fractional frequency accuracy of $10^{-17}$ can be reached for this transition because frequency shifts are almost identical for neighboring vibrational states, and several magic wavelengths that eliminate differential light shifts are known~\cite{Barontini2022Short}. The absolute frequency can be determined by referencing to a local Sr lattice clock~\cite{Baynham2026}, making an outstanding frequency standard in the MIR. A comparison over time of the molecular vibrational frequency to the optical atomic frequency will also be a sensitive probe for variations of $m_e/m_p$~\cite{Kajita2018, Barontini2022Short}. 

Locking the QCL to the trapped molecules yields a stable, local oscillator at 17.5~THz whose frequency is referenced to the standard discussed above. The stability will be determined by the linewidth of the QCL spectroscopy. In this work, the linewidth at low intensity was dominated by the laser linewidth and Doppler broadening. The laser linewidth is dominated by cavity noise between 100~Hz and 1~kHz which can be reduced with improved vibration isolation and a higher locking bandwidth. Further reduction is possible by developing a higher finesse cavity with dielectric coatings~\cite{LohnstarQuote2026}. The Doppler broadening can be eliminated using a 1D lattice. 

The main systematic frequency shift of $\ket{0,1,0} \leftrightarrow \ket{1,0,1}$ comes from the differential light shift of the two states. Over the range of intensities of interest, we calculate that at trap depth $U$ the frequency shift is $\delta f=a U+b U^2$, where $a=9.38$~Hz/$\mu$K and $b=0.117$~Hz/$\mu$K$^2$ at $\lambda=1064$~nm. The linear term is due the vibrational dependence of the scalar polarizability, while the quadratic term comes from the tensor polarizability which shifts $\ket{0,1,0}$ but not $\ket{1,0,1}$. For our trap depth, the shift is $\delta f=26.1$~kHz. The spread of frequencies is smaller than the shift by the ratio $T/U \approx 0.1$. The shift can be measured at various $U$ and extrapolated to zero. The spread can be lowered by adiabatically reducing the trap depth, but sets a practical limit of about 1~kHz to the linewidth that can be reached for this transition. The corresponding differential light shift of the Raman transition, $\ket{0,0}\leftrightarrow\ket{1,0}$, has only the linear term and is eliminated at a magic wavelength~\cite{Barontini2022Short}.  

Most research groups that need MIR laser light with high stability and accuracy do not have access to ultracold molecules and would like, instead, to lock the laser to absorption lines in common gases. In previous work with the same QCL, we measured the frequencies of 165 N$_2$O lines~\cite{Wang2025}. Despite being Doppler broadened, the statistical uncertainty in the frequencies of strong lines was about 20~kHz, but the accuracy was limited to 3~MHz by existing calibration standards. By using the CaF as a reference, and using sub-Doppler techniques, we could improve this accuracy by 2-3 orders of magnitude, providing a set of calibrated references with kHz-level accuracy that are easy for others to use. Similar methods to the ones demonstrated here can be applied to the rapidly expanding array of diatomic and polyatomic species cooled to ultracold temperatures~\cite{Fitch2021b, Augenbraun2023, Langen2024} providing a broad span of vibrational frequencies and a great leap in precision for frequency metrology throughout the mid-infrared.

\acknowledgments 

We dedicate this article to our late friend and colleague Joe Vagge. We thank Robbie Murray for the loan of a MIR-sensitive camera, Richard Hobson and Charles Baynham for helpful discussions about laser frequency stabilization, Olivier Pirali for fruitful discussions on MIR spectroscopy, and Carrie Weidner and the Bristol QE-CDT for their support. This work was supported by STFC and EPSRC under Grants ST/T006234/1, ST/W006197/1, ST/Y004345/1 and EP/W00299X/1. KKV gratefully acknowledges the Alexander von Humboldt Foundation for a Feodor Lynen Research Fellowship. The work in Paris was supported by CNRS, Université Sorbonne Paris Nord, Région Île-de-France in the framework of DIM QuanTiP and the the Agence Nationale de la Recherche (ANR projects DIABOLO (Grant No. ANR-25-CE57-7543) and Ultimos (Grant No. ANR-24-CE30-2959-01). Part of the work was carried out within the
Project 23FUN04 COMOMET, which is funded from the European Partnership on Metrology, co-financed
from the European Union’s Horizon Europe Research and Innovation Programme and supported in the UK
by UKRI (10130794).

\bibliography{references}

\onecolumngrid
\appendix

\section{End Matter\label{appendix}}

\twocolumngrid

\appendix
\refstepcounter{section}
\setcounter{equation}{0}
\renewcommand{\theequation}{\Alph{section}\arabic{equation}}

\begin{figure}
    \centering
    \includegraphics[width=\columnwidth]{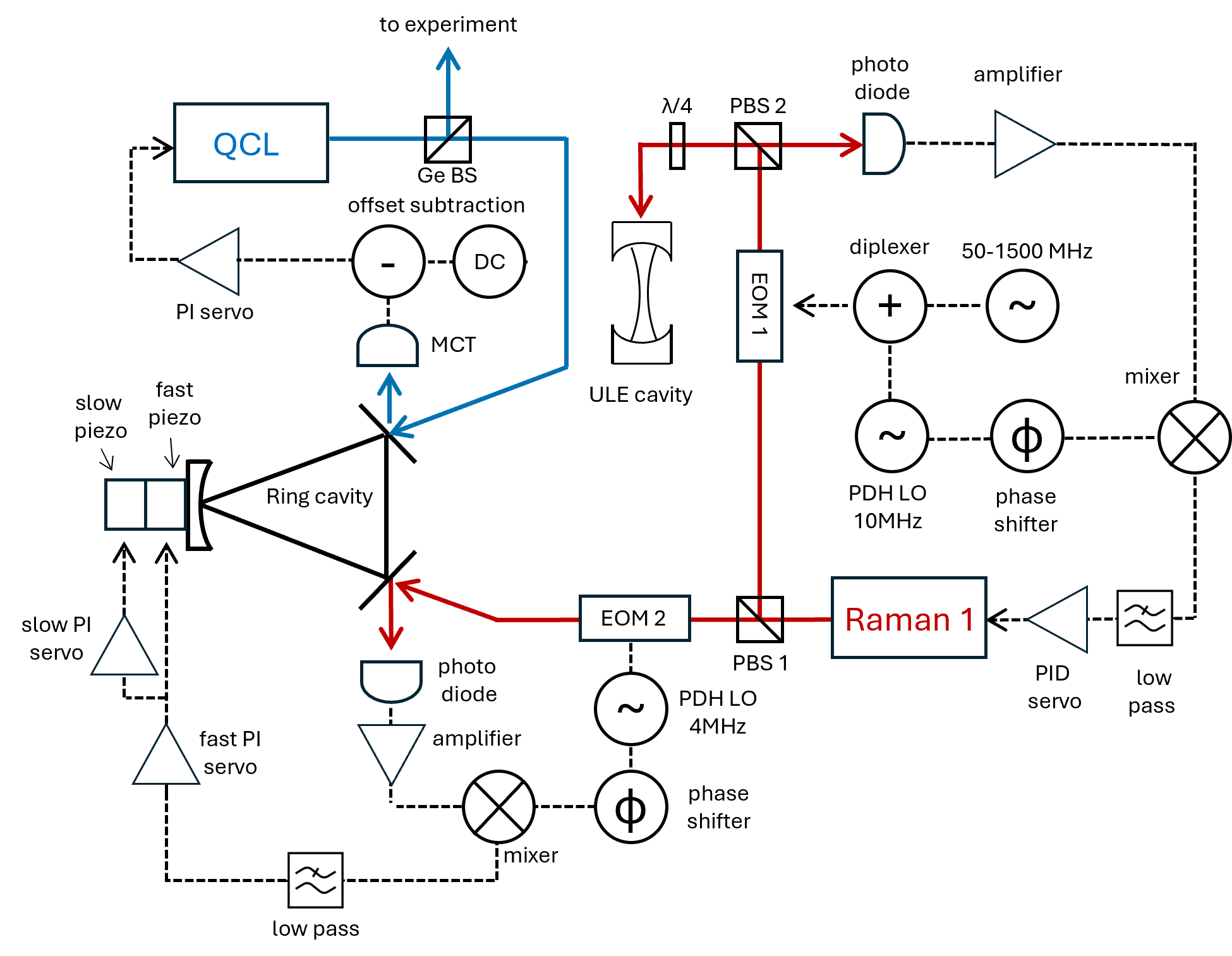}
    \caption{Schematic of the ring cavity locking scheme.
    }
    \label{fig:LockingSchemeFigure}
\end{figure}

{\it Appendix A: Locking scheme and laser linewidth}---The full locking scheme for the ring cavity is presented in Fig.~\ref{fig:LockingSchemeFigure}. The output of the 1134~nm reference laser (Raman 1) is split with a polarizing beam splitter (PBS 1). One part goes through a phase modulator (EOM 1), which is driven by the local oscillator for PDH locking at 10~MHz (PDH LO 10~MHz) as well as a frequency tunable RF synthesizer (50-1500~MHz) for offsetting the laser from the cavity mode. Both RF tones are combined on a diplexer. The light is then coupled into a high finesse ultra-stable cavity (ULE cavity) with a finesse of $3\times 10^5$. The cavity reflection is detected with an amplified photodiode and mixed down with the LO to generate the error signal, which is fed back to the laser via a PID controller. The other output of PBS 1 passes through a second phase modulator (EOM 2) which is driven by a local oscillator for PDH locking to the ring cavity at 4~MHz (PDH LO 4~MHz). The power reflected by the ring cavity is detected with a photodiode, amplified and mixed down with the LO to generate the error signal of the ring cavity PDH lock, which is then low-pass filtered and fed back via a fast and a slow integrator to the two inputs of a fast and a slow piezo stack. The QCL output is split into equal parts on a germanium beam splitter (Ge BS); one part goes to the experiment and the other part is coupled into the ring cavity. The MIR output of the ring cavity is detected with a liquid nitrogen cooled MCT detector, offset subtracted and fed back to the QCL current via a fast PI controller to lock it to the ring cavity.

To estimate the frequency stability of the QCL locked to the transfer cavity, we record the in-loop voltage error signal of the cavity locked to the reference laser (lock 1) and of the QCL locked to the cavity (lock 2). First, we scan the reference laser while recording the error signal of lock 1 together with the laser frequency measured by a high precision wavemeter. This measurement gives a free spectral range of 104~MHz and a discriminator slope for lock 1 of 4.8~MHz/V. Then, we scan the MIR frequency via the QCL current and record the cavity signal as a function of frequency, determining a discriminator slope for lock 2 of 3.9~MHz/V. Using these gradients, we convert the voltage noise of the error signals into frequency noise power spectral densities using Welch's method~\cite{Welch1967}. For both locks we also measure the contribution from the intensity noise by blocking the cavity and recording the reflected power. For the 1134~nm PDH lock this is residual amplitude noise, and for the QCL lock it is a combination of laser intensity noise, detector drifts and thermal drifts of spurious etalons which are difficult to eliminate in the MIR. 
Both intensity noise spectra are then added to the frequency noise spectra below the unity gain frequencies, which we estimate to be 300~Hz for the 1134~nm lock and 30~kHz for the QCL lock.

\begin{figure}
    \centering
    \includegraphics[width=1\linewidth]{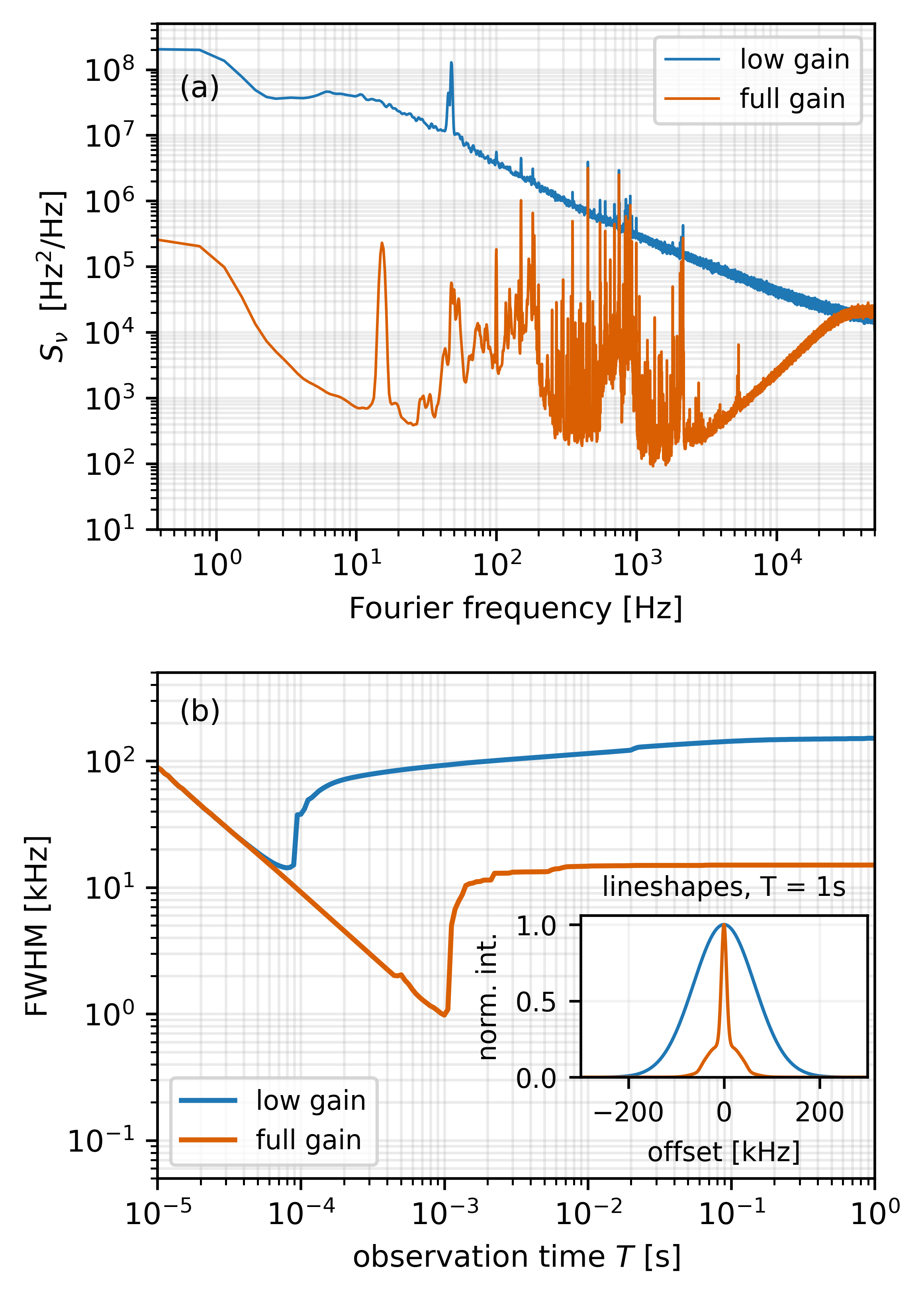}
    \caption{(a) Combined frequency noise and amplitude noise power spectral density of the in-loop error signals. (b) FWHM as a function of observation time at low and high gains, constructed from data in (a). Inset: laser lineshapes at $T=1$~s.}
    \label{fig:FreqNoisePSD}
\end{figure}

The lasers have frequencies $\nu_{\rm ref}$ and $\nu_{\rm QCL}$ and are locked to cavity modes with frequencies $\nu_{\rm ref}^{\rm cav}$ and $\nu_{\rm QCL}^{\rm cav}$. We define $r = \nu_{\rm QCL}/\nu_{\rm ref}$ and the error signals $e_{\rm ref} = \nu_{\rm ref}-\nu_{\rm ref}^{\rm cav}$ and $e_{\rm QCL} = \nu_{\rm QCL}-\nu_{\rm QCL}^{\rm cav}$. The fluctuations in the QCL frequency can be found from the fluctuations of the error signals and of the reference laser as
\begin{equation}
    \delta \nu_{\rm QCL} = \delta e_{\rm QCL} + r( \delta\nu_{\rm ref} - \delta e_{\rm ref}).
\end{equation}

Frequency fluctuations of the reference laser are negligible so the spectral density of the QCL frequency fluctuations is
\begin{equation}
S_{\rm QCL}  = S_{{\rm e}_{\rm QCL}} + r^2 S_{{\rm e}_{\rm ref}}.
\end{equation}
Here, we have neglected the cross term because $S_{{\rm e}_{\rm QCL}}$ and $S_{{\rm e}_{\rm ref}}$ are independent at $\nu<100$~Hz and $\nu>30$~kHz, while at intermediate $\nu$ the contribution of correlations is suppressed by the servo gain.

Figure \ref{fig:FreqNoisePSD}(a) shows $S_{\rm QCL}$ over a wide range of frequencies for two cases. The spectrum shown in blue was recorded with the QCL locked to the cavity at low gain, resulting in a locking bandwidth of about 10~Hz. This spectrum approximates the frequency stability of the laser under free-running conditions. The orange data is the noise spectrum recorded at high gain and a locking bandwidth of 50~kHz. These are the standard locking conditions. Under the assumption of Gaussian phase noise we calculate the normalized optical field autocorrelation function $g_1(\tau')$ for $S_{\rm QCL}$,

\begin{align}
g_1(\tau') &= e^{-\chi(\tau')}, \nonumber \\
\chi(\tau') &= \int_{1/T}^{\infty} \frac{S_{\rm QCL}(f)}{f^2}\,
\big[1-\cos(2\pi f\tau')\big]\,df.
\end{align}

The laser lineshape is the Fourier transform of $(1 - \tau'/T)g_1(\tau')$, where the pre-factor accounts for the observation window~\cite{Elliott1982, Matei2017}.  Figure \ref{fig:FreqNoisePSD}(b) shows the resulting FWHM  as a function of observation time, $T$. At small $T$, the linewidth is Fourier limited, while at longer $T$ the linewidth at high gain is dominated primarily by the noise between $10^2$ and $10^3$ Hz.  The inset shows the reconstructed lineshapes for  high and low gain at $T=1$~s. At high gain, the lineshape has a narrow central feature with a FWHM of 15.2~kHz and a broader lower-amplitude pedestal. Under nearly free-running conditions, the lineshape is Lorentzian with FWHM of 167~kHz.

\refstepcounter{section}
\setcounter{equation}{0}
\renewcommand{\theequation}{\Alph{section}\arabic{equation}}
{\it Appendix B: Mid-infrared cavity}---The MIR cavity is a ring cavity built from a curved gold mirror with focal length 1500~mm, a flat Ni-coated ZnSe mirror and a flat Ni-coated UV fused silica mirror. We choose a ring cavity to avoid optical feedback to the QCL. The flat mirrors are marketed as reflective neutral density filters with optical density 1.0 and 1.5 respectively. The ZnSe mirror is the MIR input coupler. With the MIR light incident on the Ni-coated surface at $10^\circ$ angle of incidence, we measure a transmission of $\sim$1.5\% and a reflection of $\sim$80\%. 

\begin{figure*}[t]
    \centering
    \includegraphics[width=1.8\columnwidth]{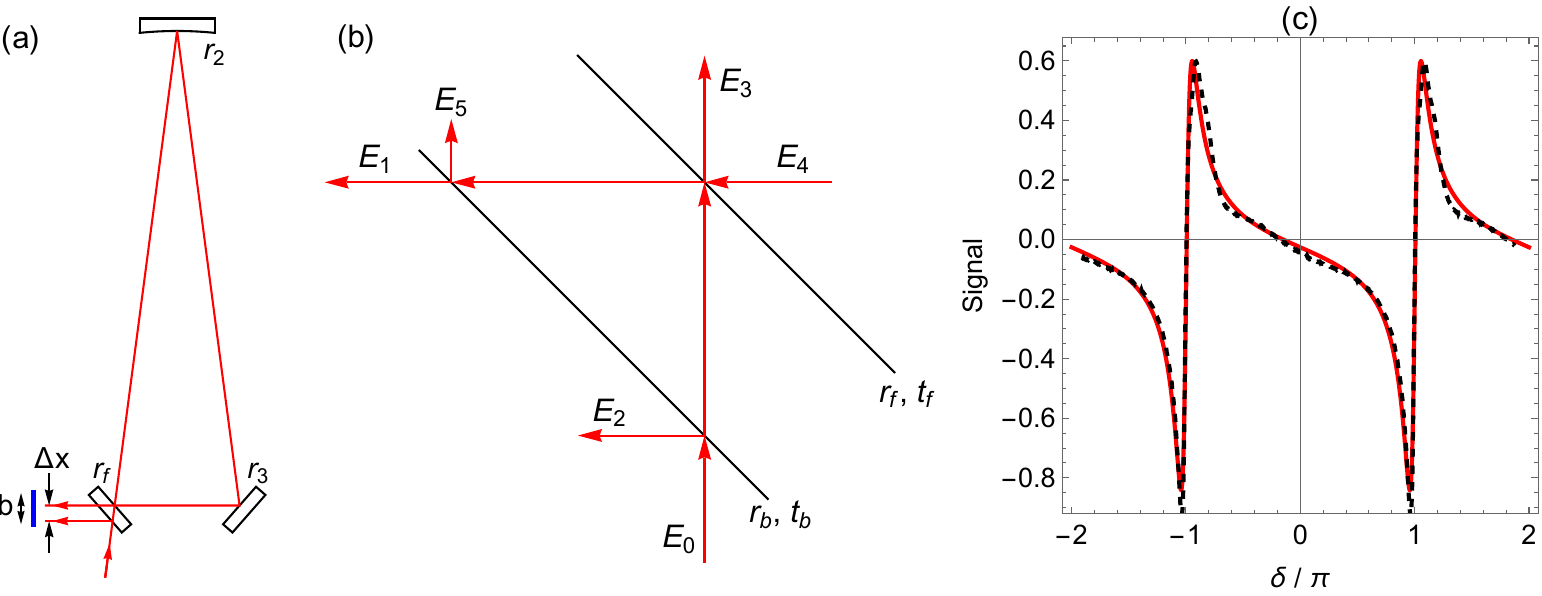}
    \caption{(a) Illustration of the ring cavity, showing some of the parameters defined in the text. Red: path around cavity. Blue: detector measuring cavity signal. (b) Light fields transmitted and reflected at the surfaces of the input coupler. (c) Dashed black line: Measured reflected signal versus laser frequency for our cavity. Solid red: Fit to the model described by Eq.~(\ref{eq:P}) with $E_1$ and $E_2$ given by Eqs.~(\ref{eq:E1}) and (\ref{eq:E2}), $f_{ij}$ given by Eqs.~(\ref{eq:f11})-(\ref{eq:f}) and fixed parameters $x_1=-\sigma/2$, $\Delta x = b = \sigma$. 
    }
    \label{fig:cavity_model}
\end{figure*}

To understand the lineshape observed in reflection from the cavity, we find it necessary to consider the reflections from both surfaces of the input coupler. Figure \ref{fig:cavity_model}(a) shows a drawing of the cavity, and Fig.~\ref{fig:cavity_model}(b) shows our model of what happens at the input coupler. The laser is incident onto the ``back'' surface which has electric field reflection and transmission coefficients $r_{\rm b}$ and $t_{\rm b}$. The light circulating in the cavity is incident on the ``front'' surface whose coefficients are $r_{\rm f}$ and $t_{\rm f}$. These coefficients are assumed to be real and to satisfy $r_{\rm f}^2+t_{\rm f}^2=1$ and $r_{\rm b}^2+t_{\rm b}^2=1$. The other two mirrors that form the cavity have reflection coefficients $r_2$ and $r_3$, and we suppose there is some round-trip attenuation $a$. The incident field is $E_0$, and the field reflected towards the detector from the front (back) surface is $E_{1(2)}$. The field transmitted into the cavity is $E_3$ and the field returning to the input coupler after a trip around the cavity is $E_4$. The light reflected at the front surface may also generate further internal reflections, labeled $E_5$, which may also be launched into the cavity. We assume this to be negligible, either because these subsequent internal reflections are too weak to matter and/or because the beam walk-off is large enough that these do not contribute to the detected signal. The phase change of the light in traveling from one surface of the input coupler to the other is $\alpha$ and the phase change in travelling around the cavity is $\delta$. Recalling the $\pi$ phase change on reflection from the air side, the fields are related as
\begin{equation}
    \begin{pmatrix}
E_1 \\ E_2 \\ E_3 \\ E_5
\end{pmatrix}
=
\begin{pmatrix}
t_{\rm b}^2 r_{\rm f} e^{2i\alpha} & t_{\rm b}t_{\rm f} e^{i\alpha}\\
-r_{\rm b}& 0 \\
t_{\rm b} t_{\rm f} e^{i \alpha} & -r_{\rm f}\\
t_{\rm b} r_{\rm f} r_{\rm b} e^{2i\alpha} & t_{\rm f}r_{\rm b}e^{i\alpha}
\end{pmatrix}
\begin{pmatrix}
E_0 \\ E_4
\end{pmatrix}
\end{equation}
The matrix $U$ that appears in this equation satisfies $U^{\dagger}U=I_2$. We also have $E_4=(\rho/r_{\rm f})e^{i\delta}E_3$ where $\rho=a r_{\rm f}r_2 r_3$ is the amplitude change for one complete trip around the cavity, accounting for absorption $a$ and reflection coefficients $r_{2,3}$ of the other two mirrors. It follows that
\begin{align}
    E_3 &= t_{\rm b} t_{\rm f} \frac{e^{i\alpha}}{1+\rho e^{i\delta}}E_0, \label{eq:E3}\\
    E_4 &= \frac{t_{\rm b}t_{\rm f}}{r_{\rm f}} \frac{\rho e^{i\delta}e^{i\alpha}}{1+\rho e^{i\delta}}E_0, \label{Eq:e4} \\
    E_1 &= t_{\rm b}^2 r_{\rm f}\left( 1 + \frac{t_{\rm f}^2}{r_{\rm f}^2}\frac{\rho e^{i\delta}}{1+\rho e^{i\delta}}\right)e^{2i\alpha}E_0, \label{eq:E1} \\
    E_2 &= -r_{\rm b} E_0 \label{eq:E2}.
\end{align}

The unusual cavity lineshape arises from partial interference between $E_1$ and $E_2$ over the limited size of the detector, so we need to include the spatial dependence of the light fields. We write the fields as $E_i({\bf r}) = E_i u_i({\bf r})$ where the $E_i$ are the complex amplitudes satisfying the relationships above, and the $u_i({\bf r})$ are real spatial functions satisfying $\int u_i({\bf r})\,d^2{\bf r}=1$. The detected power is
\begin{align}
    P &= \int_{A_{\rm det}}|E_1 u_1({\bf r})+E_2 u_2({\bf r})|^2\,d^2{\bf r} \nonumber \\
    &= |E_1|^2 f_{11} + |E_2|^2 f_{22} + (E_1^*E_2 + E_1 E_2^*)f_{12},
    \label{eq:P}
\end{align}
where the integral is over the area of the detector and
\begin{equation}
    f_{ij}=\int_{A_{\rm det}} u_i({\bf r})u_j({\bf r})\,d^2{\bf r}.
\end{equation}
We consider a square detector of side length $b$ centered at the origin, and circular Gaussian beams displaced in the $x$ direction by $x_1$ and $x_1+\Delta x$. The standard deviation of the field distribution is $\sigma$. This gives
\begin{align}
    &f_{11} = f(x_1), \label{eq:f11}\\
    &f_{22} = f(x_1+\Delta x), \label{eq:f22}\\
    &f_{12} = e^{-\Delta x^2/(4\sigma^2)}f(x_1 + \tfrac{\Delta x}{2}),\label{eq:f12}
\end{align}
where
\begin{equation}
    f(x') \!=\! \frac{1}{2} \erf\!\left(\!{\frac{b}{2\sigma}}\!\right)\!\left[\!\erf\left(\! \frac{b\!-\!2x'}{2\sigma}\!\right)\!+\! \erf\left(\! \frac{b\!+\!2x'}{2\sigma}\!\right)\!\right].
    \label{eq:f}
\end{equation}

Figure \ref{fig:cavity_model}(c) shows the measured cavity lineshape in the MIR, together with a fit to the model described above. Here, we have fixed $x_1=-\sigma/2$ and $\Delta x = b = \sigma$. The best fit parameters are $r_{\rm b}^2=0.15$, $r_{\rm f}^2 =0.83$, $\rho^2=0.74$ and $\alpha=2.84$~rad.

\refstepcounter{section}
\setcounter{equation}{0}
\renewcommand{\theequation}{\Alph{section}\arabic{equation}}
{\it Appendix C: Lineshape of vibrational transitions}---Since the laser light is applied for time $\tau$ much longer than the coherence time of the laser, coherences will have decayed away and we can model the lineshape using a two-level rate model. The transition frequency is at $\omega_0$, the natural lienwidth of the transtion is $\Gamma$, the central frequency of the laser is $\omega_L$, the total laser intensity is $I$ and the laser intensity between frequencies $\omega$ and $\omega+d\omega$ is $\rho(\omega)\,d\omega$. The excitation rate is
\begin{equation}
R=\int\frac{\Gamma}{2}\frac{1}{1+4(\Delta+\delta)^2/\Gamma^2}\frac{\rho(\omega_{\rm L}+\delta)}{I_{\rm sat}} \,d\delta,
\end{equation}
where $\Delta=\omega_L-\omega_0$, $\delta=\omega-\omega_L$, and $I_{\rm sat}=\pi h c \Gamma/(3\lambda^3)$ is the usual two-level saturation intensity. Assuming that the laser has a Lorentzian lineshape with a FWHM of $\Gamma_{\rm L}\gg \Gamma$, we obtain
\begin{equation}
    R = \frac{\Gamma}{2} \frac{s}{1+4\Delta^2/\Gamma_{\rm L}^2},
\end{equation}
with $s=\frac{I}{I_{\rm sat}}\frac{\Gamma}{\Gamma_{\rm L}}$. The excited-state population at time $\tau$ is
\begin{equation}
    N_{\rm e}\!=\!\frac{1}{2}\frac{s}{1\!+\!s\!+\!4\Delta^2/\Gamma_{\rm L}^2}\left[1\!-\!\exp\left(-\Gamma\tau\left(1\!+\!\frac{s}{1\!+\!4\Delta^2/\Gamma_{\rm L}^2}\right)\right)\right].
    \label{eq:rate_model_lineshape}
\end{equation}
When $s\gg 1$, which holds for our experiments, the pre-factor in Eq.~(\ref{eq:rate_model_lineshape}) is approximately $1/2$ and the FWHM of the lineshape is
\begin{equation}
    \Delta \omega = \Gamma_{\rm L}\,\sqrt{\frac{s \Gamma \tau}{\ln\!\left(\frac{2}{1+e^{-s \Gamma \tau}}\right)} - 1}.
    \label{eq:fwhm}
\end{equation}
Using the dipole moment data given in \cite{Pelegrini2005}, we estimate $\Gamma=4.3$~s$^{-1}$ and $I_{\rm sat}=0.18$~nW~m$^{-2}$.

\end{document}